\documentclass[aps,prb,reprint,showpacs,amsmath,amssymb,superscriptaddress]
{revtex4-1}
\usepackage{graphicx}

\begin{document}

\title{Different routes to pressure-induced volume collapse transitions in gadolinium and terbium metals }

\author{G.~Fabbris}
\affiliation {Advanced Photon Source, Argonne National Laboratory, Argonne, IL 
60439, USA}
\affiliation {Department of Physics, Washington University, St. Louis, MO 
63130, USA}

\author{T.~Matsuoka}
\affiliation {Department of Physics, Washington University, St. Louis, MO 
63130, USA}
\affiliation{KYOKUGEN, Osaka University, 1-3 Machikaneyama, Toyonaka, Osaka 
560-8531, Japan}
\affiliation{Department of Material Science and Technology, Gifu University, 
1-1 Yanagido, Gifu 501-1193, Japan}

\author{J.~Lim}
\affiliation {Department of Physics, Washington University, St. Louis, MO 
63130, USA}

\author{J.~R.~L.~Mardegan}
\affiliation {Advanced Photon Source, Argonne National Laboratory, Argonne, IL 
60439, USA}
\affiliation{Instituto de F\'isica Gleb Wataghin, UNICAMP, Campinas, SP 
13083-859, Brazil}

\author{K.~Shimizu}
\affiliation{KYOKUGEN, Osaka University, 1-3 Machikaneyama, Toyonaka, Osaka 
560-8531, Japan}

\author{D.~Haskel}
\affiliation {Advanced Photon Source, Argonne National Laboratory, Argonne, IL 
60439, USA}

\author{J.~S.~Schilling}
\email[Corresponding author: ]{jss@wuphys.wustl.edu}
\affiliation {Department of Physics, Washington University, St. Louis, MO 
63130, USA}

\begin{abstract}
\ \ \ The sudden decrease in molar volume exhibited by most lanthanides under 
high pressure is often attributed to changes in the degree of localization of their 4\textit{f}-electrons. We give evidence, based on electrical resistivity measurements of dilute Y(Gd) and Y(Tb) alloys to 120 GPa, that the volume
collapse transitions in Gd and Tb metals have different origins, despite their being neighbors in the periodic table. Remarkably, the change under pressure in
the magnetic state of isolated Pr or Tb impurity ions in the nonmagnetic Y host appears to closely mirror corresponding changes in pure Pr or Tb metals. The collapse in Tb appears to be driven by an enhanced negative exchange interaction between 4\textit{f} and conduction electrons under pressure (Kondo resonance) which, in the case of Y(Tb), dramatically alters the superconducting properties of the Y host, much like previously found for Y(Pr). In Gd our resistivity measurements suggest that a Kondo resonance is not the main driver for its volume collapse. X-ray absorption and emission spectroscopies clearly show that 4\textit{f} local moments remain largely intact across both volume collapse transitions ruling out 4\textit{f} band formation (delocalization) and valence transition models as possible drivers. The results highlight the richness of behavior behind the volume collapse transition in lanthanides and demonstrate the stability of the 4$f$ level against band formation to extreme pressure.
\end{abstract}

\maketitle

\section{Introduction}

One of the central questions in the magnetism of solids is whether the
electrons responsible for the magnetic phenomena are localized or itinerant
in nature. This dual character emerges in actinides, where the 5$f$ level is
close to a localized-itinerant boundary, leading to a large diversity of
physical properties and crystal structures.\cite{Moore2009} In lanthanides
the 4$f$ level is atomic-like at ambient pressure so that its contribution
to the material properties only occurs through interaction with the
conduction electrons. Despite the significant amount of work devoted to 4$f$
and 5$f$ electron systems over many years, the theoretical treatment of
these levels is still very challenging. Recent advances in dynamical mean
field theory have been encouraging, \cite{Zolfl2001, McMahan2003, 
McMahan2005, Lipp2012, Lanata2013} but the agreement with experiment is still
incomplete. In analogy to actinides, the localized character of the
lanthanide 4$f$ level is expected to change under sufficient pressure. \cite{Johansson1974} In particular, a sudden pressure-induced drop in the molar 
volume, commonly termed \textquotedblleft volume collapse\textquotedblright, 
has been observed in Ce (16\% volume collapse at 0.7~GPa),
\cite{Lawson1949,Lipp2008} Pr (9.1\% at 21~GPa),\cite{Cunningham2005} Eu 
(3\% at 12~GPa),\cite{Takemura1985,Bi2011} Gd (5\%  at~59~GPa),
\cite{Hua1998,Errandonea2007} Tb (5\% at 53~GPa),\cite{Cunningham2007} Dy 
(6\% at 73~GPa),\cite{Patterson2004} Ho (3\% at 103~GPa),\cite{Vohra2008} Tm 
(1.5\% at~120~GPa),\cite{Montgomery2011} and Lu (5\% at~90~GPa).
\cite{Chesnut1998} Such volume collapses are still the subject of debate,\cite{McMahan2003,Rueff2006,Maddox2006, Maddox2006a,Yoo2008,Krisch2011,Loa2012,Lipp2012,Lanata2013} 
although widely thought to result from changes in the degree of 4\textit{f} 
localization.

Here we focus on the volume collapse phenomena because the sudden and often sizeable change in the molar volume, and accompanying changes in electronic and 
magnetic properties, should facilitate the identification of its origin. In 
addition, except for Ce, the volume collapse is accompanied by a transition to 
a lower symmetry crystal structure also found in light actinides with 
itinerant 5$f$ electrons, suggesting that a fundamental change in the 
character of the 4\textit{f} electrons, perhaps  from local to itinerant, may 
take place.

Three models often invoked to describe pressure-induced volume-collapse
phenomena in the lanthanides are: (1) \textit{valence transition model}, \cite%
{Ramirez1971} where a 4\textit{f} electron is transferred into the \textit{%
spd}-electron conduction band causing a sudden reduction in the ionic radius
and enhanced metallic binding; (2) \textit{Mott-Hubbard model}, \cite%
{Johansson1974} where the 4\textit{f} states undergo a local-to-itinerant
transition, the 4\textit{f} electrons making a significant contribution to
crystalline binding; and, (3) \textit{Kondo volume collapse model}, \cite%
{Allen1982} where the approach of the localized 4\textit{f} level to the
Fermi energy under pressure leads to a sharp increase in the Kondo
temperature \textit{T$\mathrm{_{K}}$}. In all three models the 4\textit{f}
electrons play a critical role.

It is important to note, however, that 4\textit{f-}electron involvement is
not required for a volume collapse to occur. The transition metal elements Y
and Sc lack 4\textit{f} electrons, but display volume collapses of 3\% (at
99~GPa \cite{Samudrala2012}) and 4\% (at 140~GPa \cite{Akahama2005}),
respectively; both are trivalent with conduction electrons whose \textit{spd}%
-character closely matches that of the trivalent lanthanides. In fact, a
volume collapse is observed in many elements and compounds devoid of 4%
\textit{f} electrons. \cite{Louie1974,Luo1994,Lazicki2005,Xiao2010} The
volume collapse in Y and Sc is likely promoted by the ubiquitous $%
s\rightarrow d$ charge transfer under pressure whereby the number of \textit{%
d} electrons in the conduction band $n_{d}$ increases. In fact, across the
entire lanthanide series it has been shown that the variation in $n_{d}$
plays the dominant role in determining the crystal structure at both ambient
and high pressure. \cite{Duthie1977} For this reason it is important to
realize that simple $s\rightarrow d$ charge transfer must also be considered
as a viable model for pressure-induced volume collapse in \textit{all}
lanthanides, whether they contain 4\textit{f} electrons or not.

The isostructural $\gamma \rightarrow \alpha $ phase transition in Ce at
0.7~GPa exhibits the largest (16\%), and most thoroughly studied, volume
collapse of any lanthanide. \cite{Lawson1949,Lipp2008} That the 4\textit{f}
electrons play an important role in this transition is clear from the large
(80\%) and abrupt drop in the magnetic susceptibility at 0.7~GPa, \cite{MacPherson1971} thus supporting the Kondo or Mott-Hubbard models. However, 
recent results also point to the importance of lattice dynamics in Ce's volume 
collapse. \cite{Krisch2011}

For Gd and the heavier lanthanides, relatively high pressures ($>$ 50~GPa)
are required to trigger the volume collapse. Due to the technical challenges
associated with experiments at these higher pressures, the volume collapse
in the heavier lanthanides has received less attention. Both Gd and Tb
display a 5\% volume collapse at, respectively, 59~GPa and 53~GPa to a
monoclinic $C$2/$m$ structure. \cite{Hua1998,Cunningham2007} The emergence
of such a low-symmetry phase typical for the light actinides is usually
attributed to the onset of 4\textit{f} binding, i.e. the Mott-Hubbard
picture. However, the same phase is found in pure Y near 1~Mbar, \cite{Samudrala2012} thus an $s\rightarrow d$ transfer scenario cannot be ruled 
out. Both theoretical \cite{Yin2006} and x-ray spectroscopic 
\cite{Maddox2006,Yoo2008} studies report that Gd's bare local moment remains 
intact through the volume collapse transition. This result is consistent with 
both the Kondo collapse picture or simple $s\rightarrow d$ charge transfer. A 
continuous increase in hybridization between 4\textit{f} and conduction 
electrons under pressure was observed in resonant inelastic x-ray scattering 
(RIXS) measurements for Gd and interpreted as evidence for a Kondo driven 
volume collapse model. \cite{Maddox2006} However, no clear correlation between the degree of 
4\textit{f} delocalization and the volume collapse transition in Gd could be 
established. Therefore, the mechanism responsible for the volume collapse in 
Gd, Tb, and the remaining heavy lanthanides remains unclear.

We examine the origin of the pressure-induced volume collapse in Gd and Tb by 
carrying out electrical resistivity measurements on dilute Y(Gd) and Y(Tb) 
magnetic alloys to pressures as high as 120 GPa. The suppression of 
superconductivity in the Y host is used to probe changes in the magnetic state 
of the Tb and Gd ions, e.g., as a result of Kondo screening. Since the 
\textit{spd}-character of conduction electrons in Y closely matches that of 
the trivalent lanthanides, experiments on these diluted Y alloys are expected 
to mimic the interaction between 4$f$ and conduction electrons in the pure Tb 
and Gd metals, provided that 4$f$-4$f$ overlap is minimal. X-ray absorption 
near edge structure (XANES) and non-resonant x-ray emission spectroscopy (XES) 
measurements carried out in Tb metal under pressure in this work, together 
with similar measurements already published for Gd, \cite{Maddox2006,Yoo2008} indeed corroborate that 
direct 4$f$-4$f$ interactions remain largely unchanged in the studied pressure 
range, as seen from the absence of changes in local moment. The XANES and XES 
measurements allow us to probe $s\rightarrow d$ charge transfer and possible 
changes in 4\textit{f} valence and local moments. For Tb our spectroscopic 
results exclude the valence transition and Mott-Hubbard scenarios and provide 
evidence that a pressure-induced $s\rightarrow d $ charge transfer takes 
place. However, the electrical resistivity measurements strongly suggest that 
the volume collapse in Tb has a significant magnetic component and is, in 
fact, triggered by the many-body Kondo resonance. For Gd, on the other hand, 
we do not observe any clear signature of a Kondo resonance in the Y(Gd) alloys 
in the vicinity of the volume collapse transition. Considering that such 
signatures are clearly observed in Y(Pr) and Y(Tb) alloys in the vicinity of 
their volume collapses, we conclude that a Kondo-driven collapse in Gd is 
unlikely. Taken together with previous x-ray spectroscopic results on Gd 
showing absence of 4$f$ band formation or loss of local moments but significant 
$s\rightarrow d$ charge-transfer, we suggest that the latter may be the 
driving force behind the volume collapse in Gd metal.

\section{Experimental}
Dilute magnetic alloys Y(0.5 at.\% Tb), Y(0.5 at.\% Gd), and Y(1 at.\% Pr) 
were prepared by argon arc-melting stoichiometric amounts of Y and dopant (Tb, 
Pr, Y - 99.9\% pure, Material Preparation Center of the Ames Laboratory; 
\cite{ames} Gd - 99.9\% pure, Alfa Aesar). The melting procedure was repeated 
several times to promote homogeneity, the weight loss being always less than 
0.1\% of total mass. No significant concentration of other impurities or 
clustering was detected by x-ray fluorescence (XRF) and x-ray absorption fine 
structure (XAFS) measurements, as detailed in the next section. The 
high-pressure DC electrical resistivity measurements were performed in a 
membrane-driven diamond-anvil cell with both standard (300~$\mu$m culet 
diameter) and beveled (350 to 180~$\mu$m culet diameter) anvils. A rhenium 
gasket was insulated using a 4:1  c-BN-epoxy mixture which also served as 
pressure medium. The ruby fluorescence technique was used to determine 
pressure in all experiments. \cite{Chijioke2005} Four-point resistance was 
measured using leads cut from a thin Pt foil. The current used was chosen to 
keep the power dissipated in the sample always below 0.5~$\mu$W. The pressure 
cell was cooled using an Oxford He flow cryostat; after the initial cooling, 
the temperature was always kept below 120~K. The sample's lateral dimension 
was $\sim$~1/3 of culet diameter; the thickness was $<$~20~$\mu$m. The small 
sample was placed on top of the Pt leads and electrical contact was made by 
pressing the sample into the leads. Further experimental details are given 
elsewhere. \cite{Shimizu2005}

X-ray absorption near edge structure (XANES) measurements were carried out on 
a Tb foil at the L$_3$ absorption edge (2\textit{p} $\rightarrow$ 5\textit{d} 
transition) at PNC/XSD (20-BM) beamline of the Advanced Photon Source (APS), 
Argonne National Laboratory (ANL). A ``symmetric" cell (Princeton University) 
was prepared with diamonds of 300~$\mu$m bevelled to 180~$\mu$m culet 
diameter. A partially perforated plus full anvil pair was used to reduce anvil 
attenuation of x-ray intensity and improve counting statistics. A rhenium 
gasket was pre-indented to 30~$\mu$m, and a sample chamber of 80~$\mu$m 
diameter was laser drilled in the center of the indentation. A small piece of 
Tb foil (Alfa Aesar, 99.9\% purity) was loaded together with ruby balls, the 
later used for pressure calibration. Neon pressure medium was  loaded using 
the COMPRESS/GSECARS system. \cite{Rivers2008} The experiment was performed at 
room temperature, and pressure applied manually using the cell screws. XANES 
was measured in transmission mode, both incident and transmitted intensities 
were detected using N$_2$ filled ion chambers. Kirkpatrick-Baez mirrors 
focused the x-ray beam to $\sim$~3x5~$\mu$m$^2$.

Non-resonant Tb L$_{\gamma}$ x-ray emission (XES) experiments were performed 
at HP-CAT (16-ID-D) beamline of the APS, ANL. A ``symmetric" cell was used 
with full diamonds of 300~$\mu$m culet diameter. Photon energy was fixed at 
11.3~keV. Data were collected using the diamond-in gasket-out geometry, thus 
an x-ray transparent Be gasket, pre-indented to 50~$\mu$m, was used. The 
center of the gasket was replaced with a pressed c-BN/Epoxy insert. A 
100~$\mu$m hole was carefully laser drilled in the center of the insert, and 
used as sample chamber. The cell was loaded with Tb foil (Alfa Aesar, 99.9\% 
purity) and ruby balls; Si oil was used as pressure media. The experiment was 
performed at room temperature, and pressure applied manually using the cell 
screws. XES data were measured using a scintillator detector coupled to a Si 
(444) analyzer. The data were normalized by the incident beam intensity which 
was detected with a N$_2$ filled ion chamber. Kirkpatrick-Baez mirrors focused 
the x-ray beam to $\sim$~40x60~$\mu$m$^2$.

\begin{figure}[b]
\includegraphics[width = 8 cm]{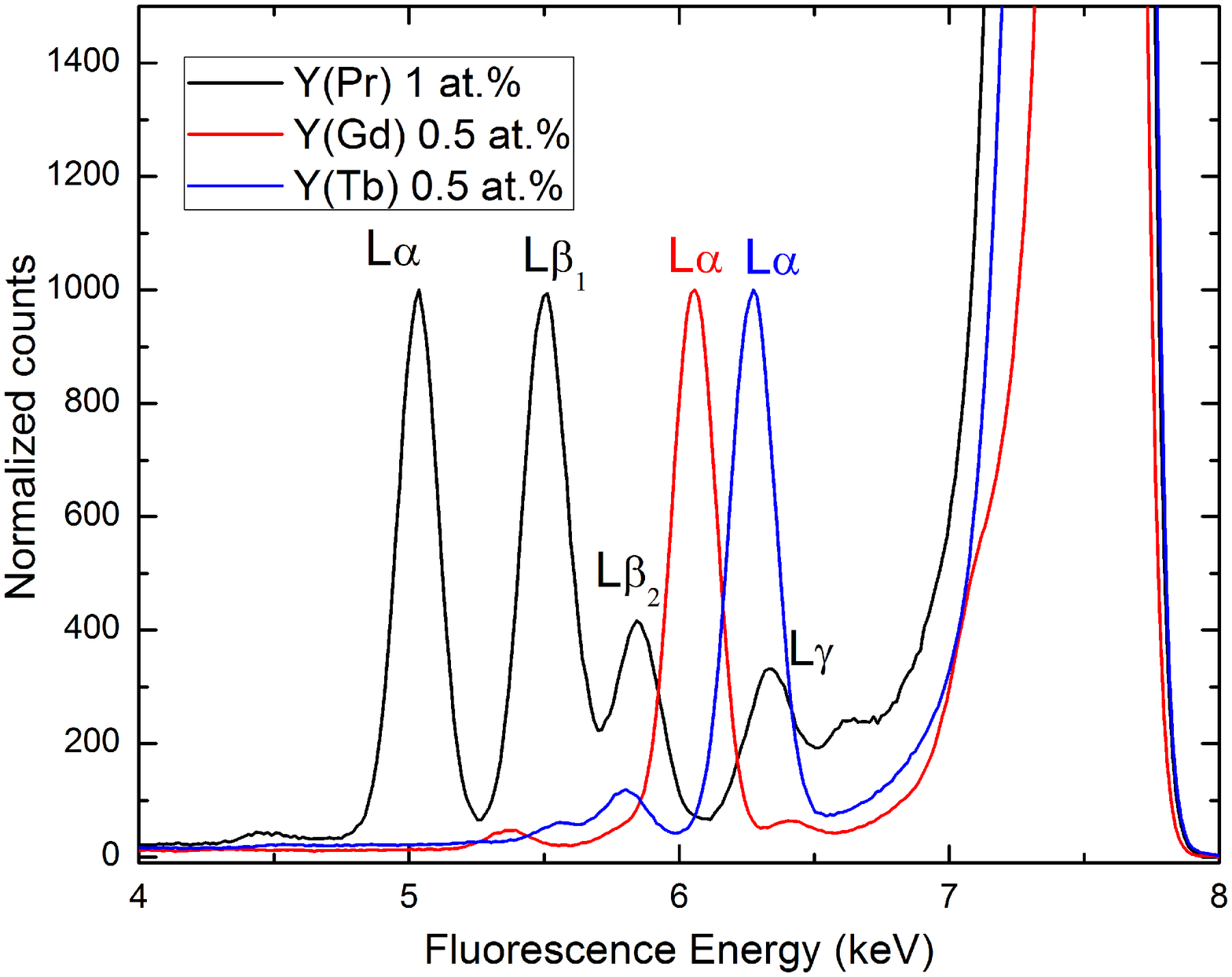} \caption{\label{fluo}(color 
online) X-ray fluorescence spectra. Data was taken at an incidence energy of 
7.55~keV and normalized to the L$_{\alpha}$ peak intensity.}
\end{figure}

\section{Sample characterization}

XRF measurements were carried out on the three Y(RE) (RE = Pr, Gd, 
Tb) alloys in order to verify the dopant content and check for unwanted 
impurities. The measurements were performed at the 4-ID-D beamline of the APS, 
ANL. An incident photon energy of 7.55~keV was used and fluorescence photons 
were collected using a four-element silicon drift energy dispersive detector 
in a normal incidence geometry. Measurements were performed at room 
temperature. Fig. \ref{fluo} shows the data normalized so that the 
L$_{\alpha}$ peak has 1000 counts. The incident photon energy lies between the 
L$_3$ and L$_2$ edges for Gd and Tb, but above L$_2$ for Pr, thus extra 
fluorescence lines appear in the latter's spectrum. This result confirm the 
presence of dopants and absence of significant presence of other impurities.

\begin{figure} [b]
\includegraphics[width = 7.5 cm]{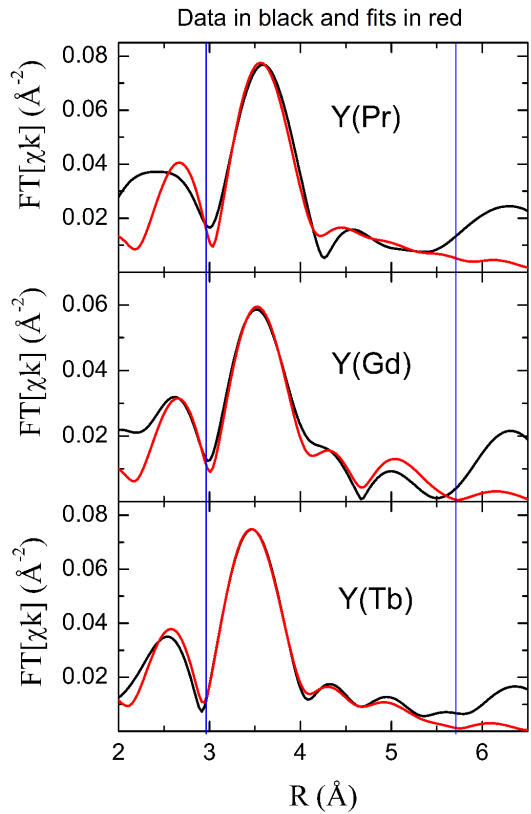} \caption{\label{exafs}(color 
online) Fourier transform of the XAFS data collected at the L$_3$ edge of Pr, Gd 
and Tb in their respective alloys}
\end{figure}

Despite the similar ionic radii of Y and dopants, the absence of dopant 
clustering has to be verified as such clustering could lead to magnetically 
ordered islands inside the Y host and potentially affect the value of 
superconducting transition temperatures, {\it T$\rm_c$}, in the Y alloys. 
XAFS measurements were carried out to probe the local environment around 
dopants. Experiments were performed at APS beamline 4-ID-D at the L$_3$ edge 
of Pr (5.964~keV), Gd (7.243~keV), and Tb (7.514~keV) using the same 
fluorescence geometry as in the XRF measurements. The photoelectron wavenumber 
was limited by the presence of the L$_2$ edge. All experiments were performed 
at room temperature. XAFS data were analysed using IFEFFIT/HORAE 
\cite{Newville2001,Ravel2005} and FEFF6 \cite{Zabinsky1995} software. The 
spectra in the k = 3-8~\AA$^{-1}$ range were truncated using a Hanning window 
and fits were perfomed in real space within r = 2.8-5.3~\AA\ up to the third 
coordination shell. XAFS spectra and fits are shown in Fig. \ref{exafs}. Fits 
with mixtures of RE-Y and RE-RE first neighbor distances were attempted. For 
the three samples the fraction of RE-RE neighbors was zero within experimental 
error ($\sim$~5\%), proving that no significant RE clustering is present. XAFS 
shows a systematic reduction in Y-RE distance in going from Pr to Tb (see 
Table \ref{dists}) as expected from the well known lanthanide contraction. 
Furthermore, the Y-RE distances are in very close agreement with the RE-RE 
distances in the pure compounds (Table \ref{dists}; we note that the first 
neighbor distances for the pure Y and RE metals are similar 
\cite{Spedding1959}). The presence of a clear lanthanide contraction in the 
Y(RE) alloys, which results from the response of outer \textit{spd} valence 
electrons to an increasingly attractive nuclear potential poorly screened by 
the additional 4\textit{f} electrons, is additional evidence for the closely 
matched character of \textit{spd} conduction electrons in the Y(RE) alloys and 
the pure RE metals. This in turn provides sensible justification for mapping 
results on the interaction between conduction electrons and local moments 
obtained from electrical resistivity measurements in the Y(RE) alloys to their 
pure RE metal counterparts.

\begin{table}[t]
\caption{\label{dists} Y-RE measured here compared to RE-RE distances obtained 
by diffraction. \cite{Spedding1959} Y-Y distance: 3.6474(7)~\AA\ .
\cite{Spedding1959}}
\begin{ruledtabular}
\begin{tabular}{c c c}
RE & XAFS (\AA) & Diffraction (\AA) \\
\hline
Pr & 3.65(2) & 3.6725(7) \\
Gd & 3.62(2) & 3.6360(9) \\
Tb & 3.59(1) & 3.6010(3) \\
\end{tabular}
\end{ruledtabular}
\end{table}

\section{Results and Discussion}

Before considering our present experimental findings on Gd and Tb, it is 
useful to first discuss earlier experiments on Ce and Pr. An almost forgotten 
strategy to test for the presence of Kondo effect phenomena is to alloy a very 
dilute concentration of the magnetic component into a superconducting host and 
see whether the pressure dependence of \textit{T$\mathrm{_{c}}$} suffers a 
characteristic \textquotedblleft sinkhole-like" suppression. \cite{Maple1976} 
Maple, Wittig and Kim \cite{Maple1969} carried out such experiments on dilute 
magnetic alloys of La(Ce) and found that \textit{T$\mathrm{_{c}}$} shows a 
dramatic \textquotedblleft Kondo-sinkhole" suppression around 0.7~GPa, close 
to the pressure where the volume collapse in pure Ce occurs. The effect of the 
very strong pair-breaking associated with the Kondo effect in dilute magnetic 
alloys has received considerable theoretical support 
\cite{Zuckermann1968,Muller-Hartmann1970,Muller-Hartmann1971,Gey1971} which 
has aided in the understanding of such complex and interesting behavior as the
reentrant superconductivity observed in La(Ce)Al$_{2}$.\cite{Maple1972}

\begin{figure}[b]
\includegraphics[width = 8 cm]{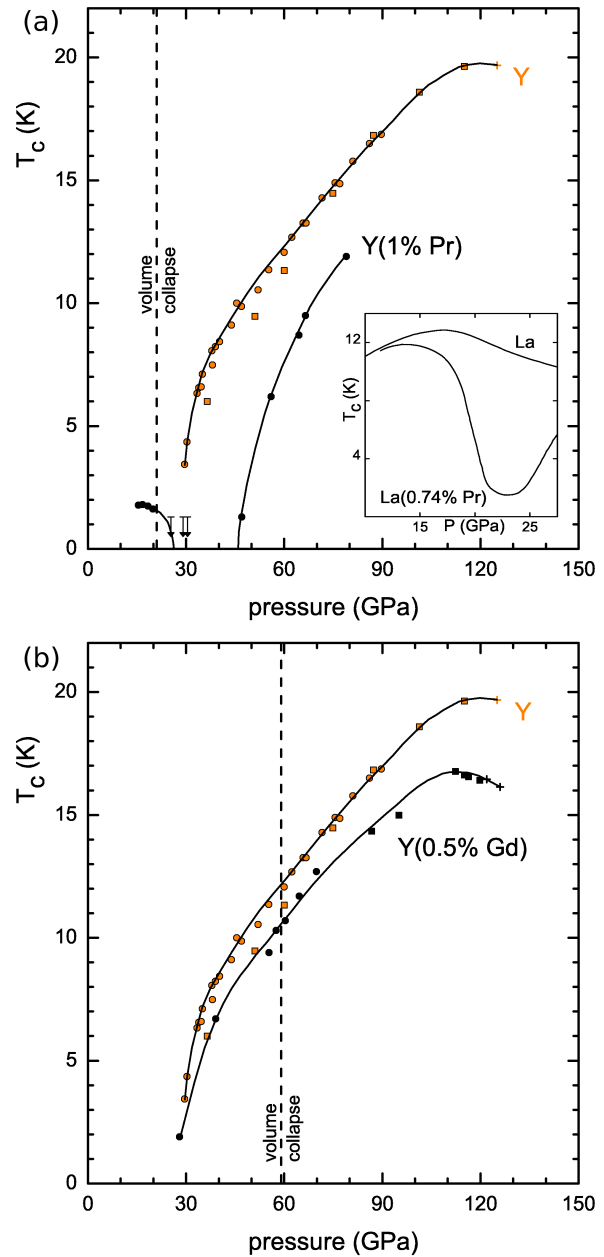} \caption{\label{yprgd}(color 
online) Pressure dependence of {\it T$\rm_c$} for (a) Y(1 at.\% Pr) and (b) 
Y(0.5 at.\% Gd) compared to that for Y. \cite{Hamlin2007} Vertical dashed 
lines show the critical pressure for the volume collapse in Pr and Gd. Inset 
in (a) shows data for La and La(0.74 at.\% Pr) adapted from Fig. 2 of Ref. 
\onlinecite{Wittig1981}.}
\end{figure}

As a second example consider Pr, which suffers a 10\% volume collapse at 
21~GPa. \cite{Cunningham2005,Mao1981} As for La(Ce), \cite{Maple1969} a marked 
suppression of \textit{T$\mathrm{_c}$} is observed in the dilute magnetic 
alloys La(Pr) \cite{Wittig1981} and Y(Pr) \cite {Wittig1982} beginning near 
21~GPa, the pressure where Pr's volume collapse occurs. These experiments were 
limited to 27~GPa so the full recovery of \textit{T$\mathrm{_c}$}(P) was not 
observed (see inset to Fig. \ref{yprgd}(a)). Fig. \ref{yprgd}(a) displays our 
recent measurements  on Y(1 at.\% Pr) which extend the previous work to much 
higher pressures. For  pressures well above 40~GPa, \textit{T$\mathrm{_c}$}(P) 
again approaches that  for pure Y. Exactly this behavior is expected as the 
Kondo temperature  \textit{T$\mathrm{_K}$} is rapidly pushed under pressure to 
values far above \textit{T$\mathrm{_c}$} where pair-breaking significantly 
weakens and the spin-compensated magnetic impurity appears nonmagnetic to the 
Y host. \cite{Zuckermann1968,Muller-Hartmann1970,Muller-Hartmann1971,Gey1971} 
XANES studies confirmed that Pr remains trivalent to 26~GPa.\cite{Roehler1995} 
XES studies on Pr metal also find no change in the bare local magnetic moment 
across the volume collapse transition,\cite{Maddox2006a} and evidence for 
4$f$-conduction electron hybridization,\cite{Bradley2012} giving strong support 
to the conclusion that this transition in Pr is Kondo-driven.

We now turn to Gd. Since the 4\textit{f}$^{7}$ orbital in Gd metal is half 
filled, its local magnetic state is the most stable of all lanthanides; Gd's 
4\textit{f}$^{7}$ level, in fact, is located $\sim $~9~eV below the Fermi 
level.\cite{Yin2006} XES experiments show no change in the local magnetic 
moment of Gd across the volume collapse transition \cite{Maddox2006,Yoo2008} 
which  excludes the 4\textit{f} local-itinerant (Mott-Hubbard) transition 
model. A small increase in the degree of 4\textit{f}$^{8}$ character observed 
in resonant L$_{\alpha }$ XES experiments\cite{Maddox2006} was interpreted as 
possible evidence for Kondo-like behavior in Gd. However, no correlation was 
found between the extent of 4\textit{f}-conduction electron hybridization 
under pressure and the occurrence of the volume collapse transition. Such 
increase in hybridization at high pressure may be interpreted as a small valence 
increase; however, no valence changes were observed with XANES,\cite{Yoo2008} 
indicating that Gd remains close to 3+ across the volume collapse, and excluding 
the valence transition model.

Further insight into the mechanism behind the volume collapse transition in
Gd is given by the \textit{T$\mathrm{_{c}}$}(P) data shown in Fig. 
\ref{yprgd}(b) for pure Y and Y(0.5 at.\% Gd). Compared to La, Y is a superior 
superconducting host for the present studies since its ionic radius closely 
matches that of the heavy lanthanides and, above 20~GPa, 
\textit{T$\mathrm{_{c}}$} increases in a simple, monotonic manner to pressures 
as high as 120~GPa (see Fig. \ref{yprgd}).\cite{Hamlin2007} In contrast to 
what is observed for La(Ce), La(Pr), and Y(Pr) alloys, no \textquotedblleft 
Kondo-sinkhole" or marked suppression of \textit{T$\mathrm{_{c}}$}(P) is 
observed in Y(Gd) at a pressure anywhere near that (59~GPa) where Gd's volume 
collapse occurs. This result suggests that the volume collapse in Gd is 
neither due to the giant Kondo resonance nor is magnetic in origin. We note 
that this is not inconsistent with RIXS results \cite{Maddox2006} which show a 
continuous increase in hybridization between 4\textit{f} and conduction 
electrons under pressure but do not establish a correlation between this 
increase and the volume collapse transition in Gd. Since robust local moments 
remain present at the collapse transition in Gd, ruling out a Mott-Hubbard 
model,\cite{Maddox2006} $s\rightarrow d$ charge transfer appears to be the 
dominant driving force for the volume collapse in Gd. This charge transfer 
causes a strong reduction in the area of the main absorption peak in Gd L$_3$ 
XANES data with pressure.\cite{Yoo2008}

\begin{figure}[b]
\includegraphics[width = 8 cm]{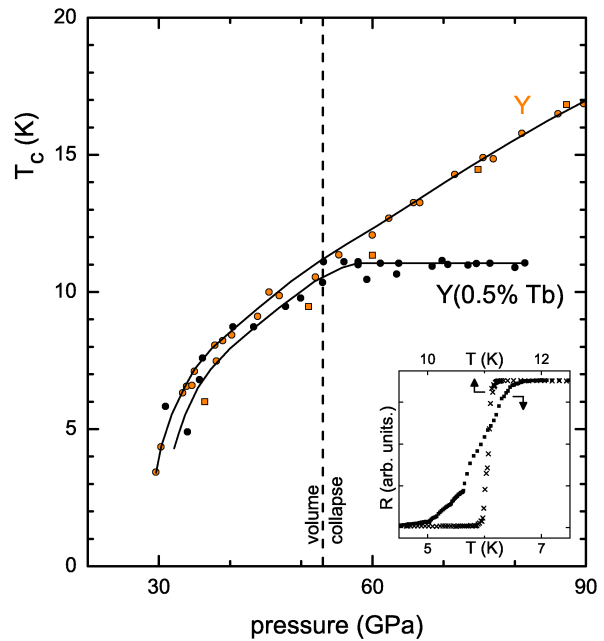} \caption{\label{ytb}(color online) 
Pressure dependence of {\it T$\rm_c$}  for Y(0.5 at.\% Tb) alloy.  Vertical 
dashed line marks pressure (53~GPa) of volume collapse in Tb. Inset shows 
resistive superconducting transition at 81.4~GPa ($\times$) is much narrower 
than that at 30.9~GPa ($\blacksquare$).}
\end{figure}

Tb is much more likely than Gd to exhibit 4\textit{f}-driven instabilities 
under pressure, since Tb's 4\textit{f}$^{8}$ level lies only $\sim $~3~eV 
below the Fermi energy.\cite{Dobrich2007} To establish if the Kondo resonance 
plays a role in Tb's volume collapse, we carried out high-pressure resistivity 
studies on Y(0.5 at.\% Tb). The \textit{T$\mathrm{_{c}}$}(P) is plotted in
Fig. \ref{ytb} for three independent runs and compared to that for pure Y. 
Beginning at about 53~GPa, a marked suppression of \textit{T$\mathrm{_{c}}$} 
is evident for the alloy with increasing pressure. Within experimental error, 
this onset pressure closely matches that (53~GPa) where the volume collapse 
occurs in Tb. The width of the resistive transition at 30.9~GPa (see inset to 
Fig. 2) arises from the pressure gradient across the sample. That this width 
becomes very narrow at 81.4~GPa is consistent with the fact that for pressures 
above 50~GPa \textit{T$\mathrm{_{c}}$} is constant. The present experiments 
thus suggest that the Kondo effect plays a role in the volume collapse 
transition of Tb metal.

\begin{figure}[b]
\includegraphics[width = 8 cm]{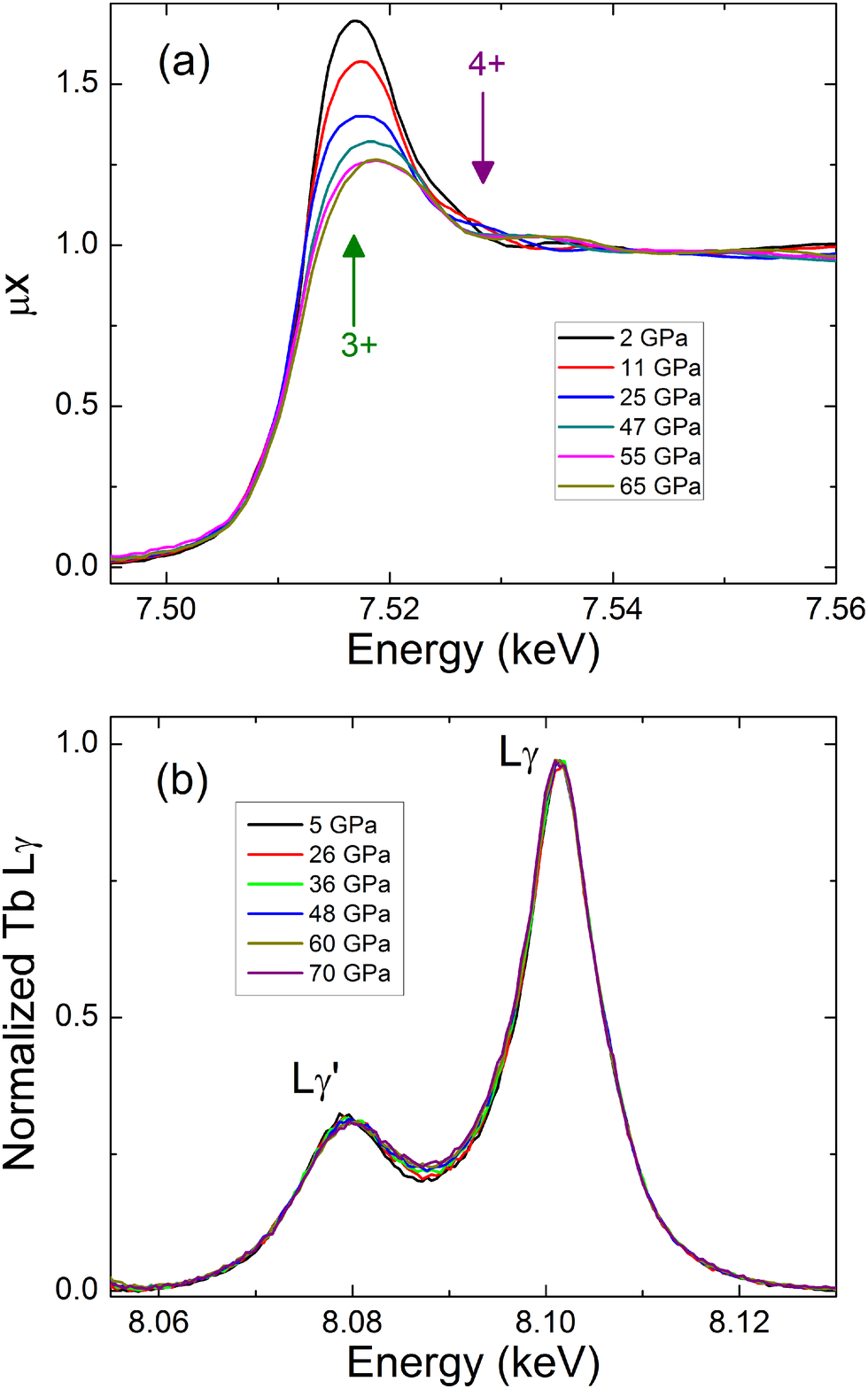} \caption{\label{tbspec}(color 
online) (a) Pressure dependence of L$\rm_3$ XANES for Tb. No 4+ or mixed 
valence state is observed. Pressure-induced reduction of peak height is direct 
measure of s$\rightarrow$d transfer. (b) Pressure dependence of L$_\gamma$ 
non-resonant XES for Tb.  A 4{\it f}\:$^8$ local-itinerant transition is not 
observed.}
\end{figure}

High-pressure L$\mathrm{_{3}}$ XANES data on Tb are presented in Fig. 
\ref{tbspec}(a). Were Tb to undergo a 4\textit{f}$^{8}$ to 4\textit{f}$^{7}$ 
valence transition under pressure, a peak would appear at the position of the 
arrow under 4+; no such transition is observed to 65~GPa. Thus a valence 
transition does not contribute to the volume collapse in Tb at 53~GPa. In Fig. 
\ref{tbspec}(a) it is clearly seen that the 3+ absorption peak is reduced in 
area with pressure. The L$\mathrm{_{3}}$ absorption edge is dominated by the 
dipolar  2$p_{3/2}\rightarrow $ 5$d$ electronic excitation; thus the area of 
the  absorption peak is directly related to the number of empty 5$d$ states. 
These  results indicate that $s\rightarrow d$ charge transfer does indeed 
occur in Tb  under pressure, suggesting that this mechanism may also 
contributes to Tb's volume collapse. However, while $s\rightarrow d$ transfer 
occurs throughout the entire pressure range measured, as also observed in 
Gd,\cite{Yoo2008} the sharp deviation in $T_{c}$(P) due to strong Kondo pair 
breaking begins at the pressure (53~GPa) where the volume collapse occurs in 
Tb, much like what is found for Ce and Pr. The Kondo resonance thus appears to 
be the main driver in Tb's volume collapse.

Changes in the character of Tb's 4\textit{f}$^{8}$ local magnetic moment can
be studied using XES. The XES L$_{\gamma}$ line is shown in Fig. \ref{tbspec}(b) 
at various pressures up to 70~GPa. In this experiment a 2$p_{1/2}$ electron is 
excited using high-energy x-ray photons. The hole is then filled by a 4$d_{3/2}$ 
electron, and the energy of the resulting L$_{\gamma}$ x-ray emission is 
analyzed with an x-ray spectrometer (Fig. \ref{tbspec}(b)). The probability of 
the  4$d_{3/2}\rightarrow $ 2$p_{1/2}$ transition depends on the initial  
(2$p_{1/2}^{1}$ 4$d_{3/2}^{4}$) and final (2$p_{1/2}^{2}$ 4$d_{3/2}^{3}$) 
states. The final 4$d_{3/2}^{3}$ state is split by the exchange interaction with 
the 4\textit{f}$^{8}$ level which leads to the splitting of the  L$_{\gamma }$ 
line seen in Fig. \ref{tbspec}(b). The ratio between the peak intensities is 
known to be related to the total angular momentum of the 4\textit{f} 
state.\cite{Jouda1997,Lipp2012,Maddox2006} No significant change in the spectrum 
is observed throughout the pressure range measured, and, in particular, no 
discontinuous change in the data is observed when the collapsed phase is 
reached. The near constancy of the L$_{\gamma}$ splitting gives direct evidence 
that the local character of the 4\textit{f}$^{8}$ orbital is preserved up to 
75~GPa, thus excluding the possibility that the Mott-Hubbard (4\textit{f} 
local-itinerant) mechanism contributes to the volume collapse in Tb.

\section{Summary}

In conclusion, taken together, the present resistivity and x-ray spectroscopy 
studies give evidence that the volume collapse in Tb metal at 53~GPa arises
predominantly from the Kondo many-body resonance, a conclusion we reached
earlier for Pr. Furthermore, we infer that the volume collapse in Gd is 
unlikely to be Kondo-driven, a conclusion based on the absence of any 
measurable effect on the $T_c$ of Y(Gd) alloys anywhere near the volume 
collapse transition. We postulate that simple $s\rightarrow d$ charge transfer 
is the main driver for the volume collapse in Gd, thus emphasizing the 
importance of considering this model as a viable explanation for volume 
collapse phenomena in 4$f$ and 5$f$ electron systems.

It may seem remarkable that the changes in the magnetic properties under 
pressure of very dilute Pr or Tb impurity ions in a nonmagnetic, 
superconducting host (Y) so closely parallel the corresponding changes in pure 
Pr or Tb metal, as evidenced by the volume collapse. That this is no accident 
is corroborated by very recent experiments on Y(Dy) and Dy.\cite{Lim2013} As 
remarked earlier, the closely matched character of \textit{spd} conduction 
electrons in the Y(RE) alloys and the pure RE metals provide a natural 
explanation for this behavior. It appears that the similarity in the interaction 
between these conduction electrons and local 4\textit{f} moments in the Y(RE) 
alloys and the pure RE metal counterparts, together with the absence of 
significant 4\textit{f} bonding, allow for mapping the observations from 
electrical resistivity measurements in Y(RE) alloys to their RE metal 
counterparts. The lanthanide contraction observed by XAFS in the Y(RE) alloys, 
which mimics the contraction observed in the pure RE metals, provides further 
validation for the correspondence in the electronic structure of outer valence 
electrons in the dilute RE alloys and RE metals.

The 5\textit{f} level in actinides is known to be close
to a local-itinerant transition,\cite{Moore2009} which has been expected to
emerge in the lanthanide 4\textit{f} electrons by $\sim $~1~Mbar within the
Mott-Hubbard model.\cite{Johansson1974} However, the present results together 
with previous x-ray spectroscopy work at high pressures indicate 
that the lanthanide 4$f$ level is considerably more stable than what is 
oftentimes assumed. For all lanthanides, except possibly Ce, pressures well 
beyond 1~Mbar appear to be required to render the 4\textit{f} electrons 
itinerant. This is consistent with suggestions of Yin and Pickett for Gd 
\cite{Yin2006} and also with considerations based on 
the degree of nearest-neighbor 4$f$ orbital overlap for all 
lanthanides.\cite{Schilling1984}

\begin{acknowledgments}
Research at both Washington University and the APS was supported by the
National Science Foundation through grant DMR-1104742 and by the
Carnegie/DOE Alliance Center (CDAC) through NNSA/DOE Grant No.
DE-FC52-08NA28554. Work at Argonne is supported by the US Department of
Energy (DOE), Office of Science, Office of Basic Energy Sciences, under
Contract No. DE-AC-02-06CH11357. J.~R.~L.~Mardegan was supported by FAPESP 
(SP-Brazil) under contract No. 2011/24166-0. The authors would like to 
thank Anup Gangopadhyay for assistance in sample preparation, as well as 
B\"orje Johansson and James Hamlin for critically reading the manuscript. We 
thank Steve Heald and Chengjun Sun for their support at Advanced Photon Source
(APS) 20-BM, Yuming Xiao, Paul Chow and Genevieve Boman for their support at
APS HPCAT 16-ID-D, and Sergey Tkachev for his help in using the APS GSECARS
gas loading system.
\end{acknowledgments}

\bibliography{paper_ref}

\end{document}